# Software Development Under Stringent Hardware Constraints: Do Agile Methods Have a Chance?

**Jussi Ronkainen**
VTT Technical Research Centre of Finland
P.O. Box 1100
FIN-90570 Oulu, Finland
+358 8 551 21040
Jussi.Ronkainen@vtt.fi

**Pekka Abrahamsson**
VTT Technical Research Centre of Finland
P.O. Box 1100
FIN-90570 Oulu, Finland
+358 8 551 2160
Pekka.Abrahamsson@vtt.fi

**ABSTRACT**

Agile software development methods have been suggested as useful in many situations and contexts. However, only few (if any) experiences are available regarding the use of agile methods in embedded domain where the hardware sets tight requirements for the software. This development domain is arguably far away from the agile home ground. This paper explores the possibility of using agile development techniques in this environment and defines the requirements for new agile methods targeted to facilitate the development of embedded software. The findings are based on an empirical study over a period 12 months in the development of low-level telecommunications software. We maintain that by addressing the requirements we discovered, agile methods can be successful also in the embedded software domain.

## 1  INTRODUCTION

Agile software development methods have captured the interest of academia and practitioners alike in the past few years. Common to the methods are the prospects of shorter lead-times, responsiveness to changes even late in the development cycle, and the promise of a continuous stream of functioning software releases from the very beginning on.

While many agile methods have been introduced (for an overview, see e.g. [1]) none of them are specifically targeted for the development embedded software. In fact, the characteristics that describe the ideal surroundings for an agile method to work best – its home ground (identifiable customer, co-located development, no more architecture design than immediately needed, object-oriented development environment, e.g. [2]) – describes the opposite of hardware-bound embedded software development. How, then, would agile development methods fit in a situation where the amount of code is not the primary scaling factor, but rather issues of performance, software reliability and constantly changing hardware requirements? This is especially the case when developing embedded systems in the telecommunications sector.

To date, there is not much literature or experiences available regarding the possible use of agile software development principles and methods in the domain of embedded software development. Yet, the electronics industry is the fastest growing industry sector in Europe. In Finland, for example, the production of electronics and electrical products has grown by a factor of eight in ten years.

Grenning [3] proposed using Extreme Programming [4] in the development of embedded software, but in his development the hardware was not a major player in the product development until late in the project. In the environment we studied, however, the hardware is available already at an early stage of a project, causing much change into the software development.

We base our work on an empirical study performed in a tightly hardware-bound environment where the aim was to improve the existing processes. The details of the study can be found in [5]. Drawing from this experience we analyze the prospects of using state-of-the-art agile methods in developing embedded software under tight hardware constraints. On this basis, we finally define the requirements for new agile methods targeted fit for this domain of software development.

## 2  EMBEDDED SOFTWARE IN DIGITAL SIGNAL PROCESSING APPLICATIONS: THE FOUR CHARACTERISTICS

Embedded software can be found in a wide variety of applications and the environment, requirements and constraints for different types of software in a single system vary. We focus on the specific problems in writing software that directly accesses hardware.

Our specific interest is in digital signal processing applications. Data processing in such systems typically uses general purpose microprocessors or microcontrollers, digital signal processors (DSPs) and application-specific integrated circuits (ASICs). General-purpose processors are mainly used for controlling overall system functionality and running user interface-related tasks. Digital signal processors and application-specific integrated circuits are used in performing computationally intensive signal processing tasks. DSP software allows flexibility in implementation and makes it possible to update the system through new software releases. The most intensive mathematical tasks are implemented in the ASICs. As telecommunication systems become more complex, the amount and complexity of the software required to facilitate the co-operation of these technologies grows steadily.

**Characteristics of Embedded System Development**

The development of embedded systems is characterized by the need to develop software and hardware simultaneously [6]. This concurrent work is known as co-design. In our case, this means that the DSP software and ASICs are concurrently under development, giving the software development work a unique flavor. The simultaneous development means that overall system functionality is constantly balanced between hardware and software implementation, and that a significant part of the software has to be developed in close co-operation with the concurrent hardware development. We call this software "hardware-related". The concept of co-design in such a case is illustrated in Figure 1.

The dynamics of co-design – i.e. the way it effects the concurrent software development processes, has to be understood in order to enable the use of agile software development methods.

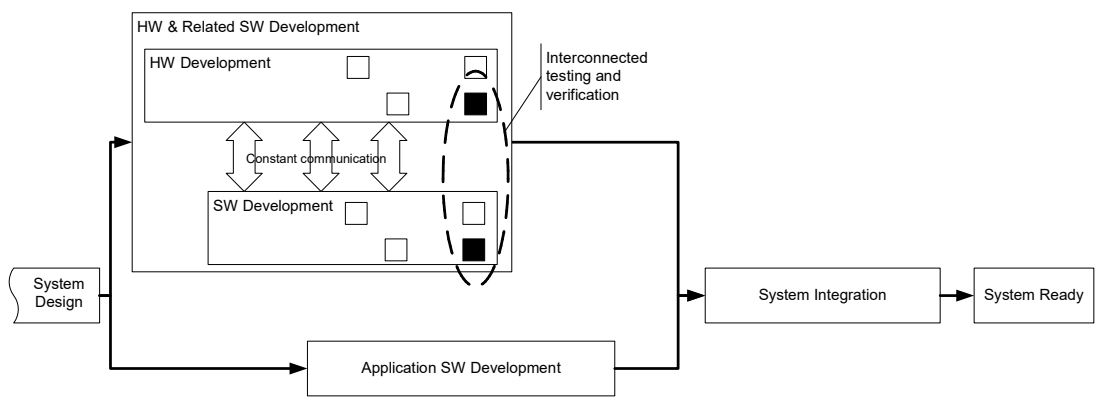

Figure 1. Co-design timeline example [5].

During our research into hardware-related software processes, we found four characteristics in hardware-related software development that rose above others from an agile software development viewpoint. We will use these characteristics later as a basis for an analysis of the specific problems that arise from closely hardware-related software work, and thus of the suitability of agile software development methods for developing such software. The characteristics and the specific effects they have on using agile methods are described in the following.

**Meeting the hard real-time requirements is the number one priority**

The environment in which the software runs imposes several strict requirements for the software. Some of the most essential requirements concern performance. Embedded systems often have to perform tasks within a time slot defined by e.g., a telecommunication standard. If failing to comply the timing and performance requirements results in a risk to further system operation or in other considerable, non-correctable errors, the real-time requirements are said to be hard [7]. In hardware-related software development, the hard real-time requirements are visible most concretely in the constant need to verify the proper co-operation of DSP software and the ASIC hardware the software drives. This causes that hardware simulators are an essential tool during DSP software development. The hardware simulators are also actively used in the development of the ASICs, binding the two development lines tightly together in terms of schedule and the use of the shared simulation resources.

The use of hardware simulation also makes it possible to make the final split between hardware and software functionality at a fairly late stage during development. From a software development viewpoint this means that the requirements for hardware-related software cannot be frozen before development work begins. Other notable technological constraints that cause changes during development are those of memory and power consumption. Therefore, the development method has to have some kind of mechanism to cope with changes in requirements during development.

A considerable deal of architecture development is practically mandatory in composing the functionality for the system. Some of the architecture emerges through experience gained during development, but preliminary architecture design cannot be avoided. Most agile methods do not encourage this. Furthermore, agile methods do not motivate rework due to lack of software performance – another commonplace activity in embedded software development. Therefore, the concept of using "the simplest solution that could possibly work" (stated, e.g., as the "YAGNI" principle in XP [4]) must be stretched somewhat.

Another key issue is refactoring. This practice of customary rearrangement and clean-up of code in order to simplify it or make it more understandable is an everyday practice in, for example, Extreme Programming. Refactoring high-speed hardware-related code is, however, hazardous. The interactions between the software and the hardware are numerous, complex and typically very sensitive to changes in timing. Therefore, thoughtless changes in code – even if the code logically remains the same – may cause slight changes in timing or other behavior, which turns into bugs that are very difficult to detect. The negative effects of refactoring can be alleviated through pervasive use of software configuration management and relentless testing, but the latter has its own problems, as we'll discuss later.

**Experimenting is part of the development**

The way the technological constraints (performance, power and memory consumption, etc.) effect code is impossible to tell exactly without hands-on experience. Therefore, the more complex the software-hardware interactions, the more the developers will experiment. This is not quite unlike the use of spike solutions in XP or prototyping in general. The difference in hardware-related software development is that the amount of code that is generated through experimenting is very significant, and much of it will evolve into actual production software.

As the development progresses, the code is required by more and more stakeholders (other software teams, hardware teams, production teams), and the effects of changes in hardware or related software ripple substantially farther than within the work of the corresponding teams. Therefore, the rigidity of software development practices has to steadily increase from what is needed in the initial, turbulent environment where changes have limited impact, to the final stages where the slightest changes have to be carefully analyzed and accepted among several stakeholders. This kind of on-the-fly adjustment of the practices is not adequately supported by current agile methods.

**High level designs and executable documentation are not sufficient**

The information transferred between the teams implementing the system is typically very specific as regards timing, bit patterns, etc. Furthermore, embedded system development requires a wide range of expertise, which means that distributed development is often a necessity. While individual teams may still reside on a shared location, the mix of different technologies involved requires communication across different teams, all of which cannot be expected to use the software as the only documentation. Therefore, information exchange cannot solely consist of face-to-face communication with source code as the only documentation. Furthermore, synchronizing the teams' work requires a certain amount of up-front design documentation.

The inability to avoid up-front documentation is an obvious challenge to fully-fledged use of agile methods. The problem of keeping the documentation up to date remains, however. Therefore, the challenge for agile methods is to provide more sophisticated methods for recognizing the required amount of documentation at a given time, rather than sticking to the idea that working software is always sufficient.

**The development is test driven by nature**

The most predominant activity in developing complex embedded systems is testing. The requirements for embedded system reliability and device autonomy are generally strict [8]. In addition to the normal software tests (unit, integration, acceptance), many tests focus on the functionality of the hardware the software drives.

Some testing concepts promoted by agile approaches (the use of regression tests, for example) are already in place in hardware-related software development. Some of the core ideas (write tests first, run every unit test at least daily) are problematic, however, in hardware-related software development. The test environment is usually different from the development environment, and memory or performance constraints often prevent installing and running every all of the test code in the testing environment at the same time. Further still, daily testing may not be possible due to the sharing of the hardware simulation resources with hardware teams.

Despite the problems, the agile approach to testing is definitely worth investigating in the realm of hardware-related software development. Specific approaches are required, however, for mitigating the problems of scaling the test software to different situations.

## 3 SPECIFIC REQUIREMENTS FOR AGILE METHODS IN EMBEDDED SOFTWARE DEVELOPMENT

Closer inspection of the characteristics of embedded system development shows that the problems faced in the turbulent software-hardware boundary are largely those the agile methods are intended to solve. In particular, constant change in requirements and the need to experiment already necessitate the use of an iterative and incremental development process. Testing is also vital in embedded software development, yet another highly encouraged practice in agile development methods. Finally, efficient and timely communication between hardware and software developers is paramount.

Table 1, based on the discussion above, puts forward four basic problems areas, their descriptions and the embedded domain specific requirements for the new agile software development methods.

**Table 1. Specific problems and requirements.**

| Problem area | Problem description | Embedded domain requirements |
|---|---|---|
| Hard real-time requirements | Up-front design and architecture work cannot be avoided.<br><br>Extensive refactoring not always feasible. | Models needed for determining specification and documentation levels. Techniques and methods needed for the development of a flexible architecture.<br><br>Refactoring integrated with a workable configuration management system that includes relevant hardware versions. System-level impact analysis methods. |
| Experimenting | Moving from prototype code to well-documented production code is a challenge. | Models needed for progressively increasing code maturity. Different scales for code in different phases of evolution. |
| Documentation | Executable documentation alone is not possible.<br><br>Gradually growing number of stakeholders requires more and more rigorous methods.<br><br>Distributed development across several teams requires more than just face-to-face communication. | Techniques needed for recognizing and managing change-prone requirements.<br><br>Ways to enable a gradual introduction of more rigid practices.<br><br>Coordination and communication methods for inter-team work. |
| Test-driven development | Added code for testing software effects system performance and hence, test results. Capacity constraints restrict the amount of test software on the system. | Test software has to be flexible in terms of size and control – only the essential for performance, more extensive for testing program logic. |

The good news for hardware-related software developers is that many of the principals proposed in the agile methods are universal enough to suit the development of complex embedded systems. The challenge, however, is that the current operationalization of these principles, i.e. the existing agile methods, do not suit to the development of hardware-related software as such.

Pervasive use of version / configuration control is one key ingredient in enabling fast-paced development work in an environment where seemingly harmless changes may cause bugs that are very difficult to locate and fix. This also has to entail relevant hardware development versions (simulation models etc.), as the functionality of software always has to be verified against the hardware, and vice versa.

Currently, existing agile methods can be most effectively utilized during the early phases of development, when even the most essential requirements may be unclear, and the availability of any working software is crucial in helping the concurrent hardware development. The key issue in adapting agile methods into usable solutions in embedded system domain is development time scalability. What is thus required is a method with the ability to scale smoothly during development to cater for the increasing need of formal communication, change management methods, and documentation.

## 4  CONCLUSIONS

We described the essential characteristics of hardware-related software development, and analyzed the suitability of agile development methods in developing the software. As we pointed out, the development of this type of software has to face many of the same problems the agile methods were created to solve. Currently available agile development methods do not, however, adequately address the specific characteristics of hardware-related software development.

Our analysis was on a very limited area of embedded software development. However, since embedded system development in general is characterized by the simultaneous development of software and hardware, the problems we described are not unique to the development of the most hardware-bound software.

Agile methods certainly do have something to offer for the development of embedded software. Therefore, in order to establish a foothold in the development of embedded systems, agile methods have to focus on the specific embedded domain requirements we set out.